\documentclass[%
 reprint,
 amsmath,amssymb,
 aps,
]{revtex4-1}

\usepackage{graphicx}
\usepackage{dcolumn}
\usepackage{bm}
\usepackage{xcolor}

\begin{document}

\preprint{APS/123-QED}

\title{High-Fidelity Simultaneous Detection of Trapped Ion Qubit Register}

\author{Liudmila A. Zhukas}
\email[]{Corresponding author: lzhukas@uw.edu}
\affiliation{Department of Physics, University of Washington, Seattle 98195, USA}
\author{Peter Svihra}
\affiliation{Department of Physics, Faculty of Nuclear Sciences and Physical Engineering, Czech Technical University, Prague 115 19, Czech Republic}
\affiliation{Department of Physics and Astronomy, 
School of Natural Sciences, 
University of Manchester, Manchester M13 9PL, United Kingdom}
\author{Andrei Nomerotski}
\affiliation{Physics Department, Brookhaven National Laboratory, Upton, NY 11973, USA
}
\author{Boris B. Blinov}
\affiliation{Department of Physics, University of Washington, Seattle 98195, USA
}

\date{\today}

\begin{abstract}
Qubit state detection is an important part of a quantum computation. As number of qubits in a quantum register increases, it is necessary to maintain high fidelity detection to accurately measure the multi-qubit state. Here we present experimental demonstration of high-fidelity detection of a multi-qubit trapped ion register with average single qubit detection error of 4.2(1.5) ppm and a 4-qubit state detection error of 17(2) ppm, limited by the decay lifetime of the qubit, using a novel single-photon-sensitive camera with fast data collection, excellent temporal and spatial resolution, and low instrumental crosstalk.
\end{abstract}

\maketitle

Trapped ions are among the most promising candidates for practical quantum computing due to a combination of unique properties, including very long coherence times, high fidelity qubit state initialization, manipulation and detection, and prospects for scaling up~\cite{Monroe13, Wineland2008}. State-dependent fluorescence is used to detect trapped ion qubit state. It relies on the existence of a cycling transition, which includes one of the qubit states (called the ``bright" state) and excludes the other (the ``dark" state)~\cite{Wineland80}. A single ion scatters a large number of photons when in the bright state, which are collected and detected. An ion in the dark state does not scatter any photons. Simple discrimination between of the number of detected photons provides single-shot measurement of the qubit state. Scaling up the trapped ion system requires counting the number of photons individually for each ion. Thus, an optical system and a photon-counting detector with sufficient spatial resolution is necessary.

Fidelity of multi-qubit state detection in a trapped ion chain depends on the integration time, photon collection efficiency, performance of the optical system, instrumental noise, and detection crosstalk. 
Single ion qubit state detection fidelity of up to 0.99971(3) has been demonstrated in $^{133}$Ba$^{+}$ using a photomultiplier tube (PMT) \cite{christensen2020}. Simultaneous detection of multiple ions requires spatially-resolving detectors. Electron-multiplying charge-coupled device (EMCCD) cameras are commonly used  \cite{Avella2016,Moreau2019}. A single $^{40}$Ca$^+$ qubit readout error as low as 0.9(3)$\times$10$^{-4}$ using an EMCCD was demonstrated, limited by the 1.168(7)~s spontaneous emission life time of the qubit \cite{Burrell2010}. Similar camera has been used to measure the state of a 53-ion qubit register \cite{zhang2017} with nearly 0.99 single-qubit detection efficiency. However, the serial interface of a CCD camera is slow. Segmented multi-anode PMTs offer fast, on-demand detection with some degree of spatial resolution. However, due to crosstalk between the PMT channels, multi-qubit state detection fidelity is lower than the product of the single-qubit fidelities. For example~\cite{Linke2017}, a single-qubit detection fidelity of 0.994 was observed in a 5-qubit system using a segmented PMT, while the 5-qubit state detection fidelity was only 0.957, which is noticeably lower than 0.970 expected from the independent error model. To lower the crosstalk, superconducting nanowire single photon detectors (SNSPDs) have been used for state detection of two ions \cite{Crain19} with average qubit state detection time of 11~$\mu$s and average fidelity of 0.99931(6). However, scaling up the number of SNSPDs to tens and hundreds is challenging.

Here we demonstrate simultaneous detection of four $^{138}$Ba$^{+}$ ion qubits. The qubit is spanned by the $6S_{1/2}$ ground state and the $5D_{5/2}$ metastable state (spontaneous emission life time $\tau=31.2(0.9)$~s~\cite{Auchter2014}) of the ions. Ion detection was done with a time-stamping, single-photon-sensitive camera Tpx3Cam~\cite{timepixcam, Nomerotski2019, zhukas023105}. The camera has a high quantum efficiency (QE) back-side illuminated optical sensor~\cite{Nomerotski2017}, bump-bonded to the Timepix3~\cite{timepix3}, an application-specific integrated circuit with 256$\times$256 pixels measuring 55$\times$55 $\mu$m$^2$ each. Electronics in each pixel processes the incoming signals to measure their time of arrival (ToA) for hits that cross a predefined threshold with 1.56~ns temporal granulation. Information about time-over-threshold (ToT), which is related to the deposited energy in each pixel, is stored together with ToA as time codes in a memory inside the pixel. The Timepix3 operation is data-driven, with pixel dead time of only 475~ns~+~ToT allowing for independent multi-hit functionality for each pixel with 80 Mpix/sec total bandwidth.

For the single photon operation, the signal is amplified using a Cricket$^{TM}$ adapter~\cite{cricket} with integrated image intensifier and relay optics to project light flashes from the intensifier output window directly on the optical sensor of the camera. 
The image intensifier is a vacuum device comprised of a photocathode followed by a micro-channel plate (MCP) and fast scintillator P47. 
The hi-QE-green photocathode in the intensifier has QE of about 20\% at 493~nm. The MCP in the intensifier had an improved detection efficiency close to 100\%.
Similar configurations of the intensified Tpx3Cam were used before for characterization of quantum networks~\cite{Ianzano2020, Nomerotski2020}, quantum target detection~\cite{Yingwen2020, Svihra2020}, single photon counting~\cite{sensors2020} and for lifetime imaging~\cite{Sen2020}.

\begin{figure}[t]
\includegraphics[width=3in]{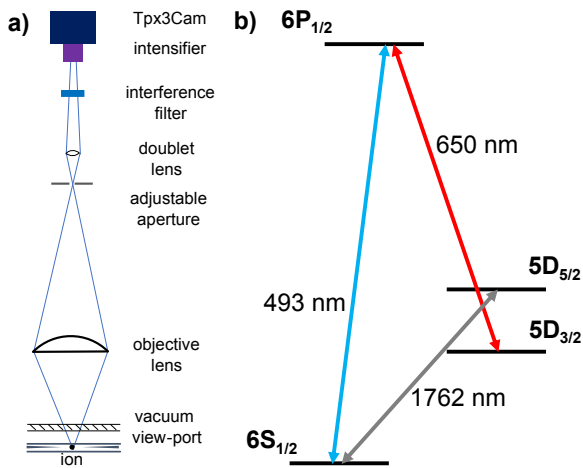}
\caption{\label{fig:epsart111}Optical setup (a) and energy level diagram of $^{138}$Ba$^{+}$ (b). Fluorescence from ions passing through the vacuum view-port is collected by the objective lens, through the adjustable aperture and further magnified by the doublet lens. An 493~nm interference filter is used to reduce the background. Finally, the 493~nm fluorescence is imaged onto the intensifier, which is connected to the Tpx3Cam camera. 493~nm and 650~nm lasers drive $6S_{1/2}-6P_{1/2}$ and $5D_{3/2}-6P_{1/2}$ cooling and repumping transitions, respectively. A 1762~nm laser couples the $5D_{5/2}$ and the $6S_{1/2}$ qubit states.}
\end{figure}

After ordering in time, pixels are grouped into ``clusters" using a recursive algorithm~\cite{tpx3cam}. Clusters are small collections of adjacent pixels within a predefined 300~ns time window. Since all hit pixels measure ToA and ToT independently and provide position information, they can be used for centroiding to determine the coordinates of single photons. ToT information is used for the weighted average, giving an estimate of the x and y coordinates for the incoming single photon. The timing of the photon is estimated by using ToA of the pixel with the largest ToT in the cluster. This ToA is then adjusted for the so-called time-walk, an effect caused by the variable pixel electronics time response, which depends on the amplitude of the input signal \cite{Turecek_2016, tpx3cam}.

The 4-ion chain of $^{138}$Ba$^{+}$ was stored in a ``five-rod" linear RF trap~\cite{Dietrich2010}. To Doppler-cool ions, $6S_{1/2}-6P_{1/2}$ transition near 493~nm was used. A 650~nm laser repumped ions from the long-lived $5D_{3/2}$ metastable state. A 1762~nm fiber laser was used to coherently drive the $6S_{1/2}-5D_{5/2}$ quadrupole transition, which is the qubit transition in this experiment \cite{Wright2016}. In this work, $5D_{5/2}$ and $6S_{1/2}$ are referred to as the dark and the bright states, respectively. The ion does not couple to the cooling/repump lasers when in the $5D_{5/2}$ state, so no fluorescence is detected; when in the $6S_{1/2}$ state, the ion scatters $\sim10^{7}$ photons/s. Relevant energy levels and transitions in $^{138}$Ba$^{+}$ are shown in  Figure \ref{fig:epsart111}(b). 

The optical system is shown schematically in Figure \ref{fig:epsart111}(a). It consists of an objective lens (50~mm Nikon lens with numerical aperture 0.20), an adjustable aperture to filter out stray light, and a secondary lens (home-built 25~mm doublet). A 493~nm interference filter suppressed the background light. The magnification of the system is approximately 45, and its collection efficiency is approximately 1.3\%.

\begin{figure}[b]
\includegraphics[width=3in]{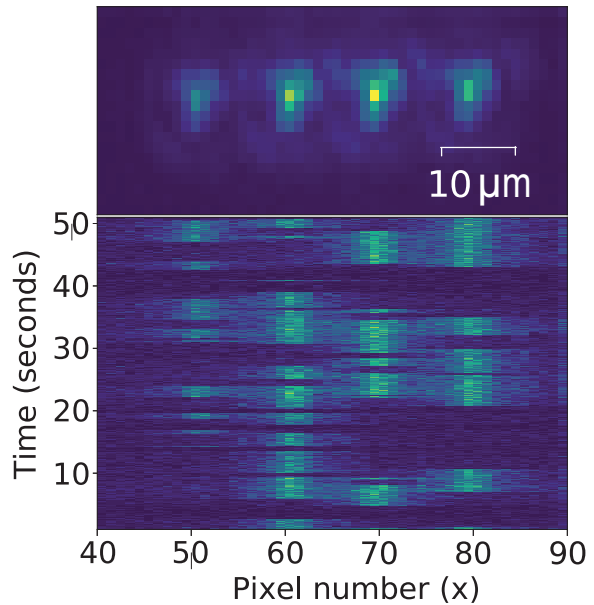}
\caption{\label{fig:epsart} Image of 4 laser-cooled $^{138}$Ba$^{+}$ ions in a linear trap taken with an intensified Tpx3Cam camera, with a typical time sequence of all 4 ions undergoing quantum jumps shown directly below. The ion separation is approximately 10$\mu m$.}
\end{figure}

We use threshold analysis method~\cite{Burrell2010, Keselman2011HighfidelitySD} to calculate the qubit state detection error. We count the number of detected photons for each ion during a set time interval called the integration time, with $n_{b}$ being the number of photons for the bright state and $n_{d}$ for the dark state. $n_{d}$ and $n_{b}$ are random variables whose probability distribution functions (PDF) are well approximated Poisson distribution. The threshold method is based on setting a specific value $n_{tr}$, such that if the number of detected photons is greater than $n_{tr}$, then the ion state is bright, while if the number is lower than $n_{tr}$, then the ion state is dark. The optimal value of $n_{tr}$ is near the intersection of dark state and bright state PDFs, where the value of the state detection error reaches its minimum. The detection error is defined as $(\epsilon_{d}+\epsilon_{b})/2$, where $\epsilon_{d}$ is the probability to misidentify a dark state as bright and $\epsilon_{b}$ is the probability to misidentify the bright state as dark.

\begin{figure}[t]
\includegraphics[width=3.3in]{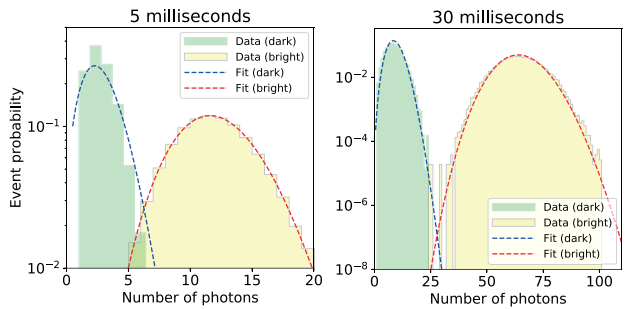}
\caption{\label{fig:epsart2} The bright and dark state histograms for the second ion from the left for two integration times. The probability to register a certain number of photons during the integration interval is plotted versus the number of photons, separately for dark and bright qubit states. The dashed curves are Poissonian fits to the data.}
\end{figure}

Data was collected with frequency and intensity of the 493~nm and 650~nm lasers set to efficiently Doppler-cool the ions. The top panel in Figure \ref{fig:epsart}(a) shows an image of four $^{138}$Ba$^{+}$ ions in the bright state. The 1762~nm laser was turned on at a low intensity, such that the ions underwent quantum jumps between the bright and dark states at a rate of approximately 1 per second or less. 
Note that due to slight misalignment of the 1762~nm laser beam the rate of quantum jumps was different for the four ions, as is evident from Figure \ref{fig:epsart}(a), where the rate is greater for the leftmost ion. This difference, however, does not affect the results presented in this work. 

Analysis was performed on several data sets of four ions undergoing the quantum jumps, amounting to 12~hours in total. We performed data post-selection to identify the state of each ion, making use of individual photon detection with reliable time of arrival. Time delays between the detected photons differ significantly between the dark and bright ion states. For this experimental setup, the average time between photons is approximately 0.5~ms for the bright state and 3.5~ms for dark state. Transitions between states can be identified by an increase or decrease of time delays between the detected photons. We set two temporal thresholds for identifying the transitions. If the time delay between two consecutive photons exceeds the upper threshold, the ion is in the dark state; if the time between consecutive photons is less than the lower threshold, the ion is in the bright state. Time delays that lie between the two thresholds do not provide enough certainty of the ion state and are excluded from analysis. A qubit transition has occurred in between the detection of two photons if the time delays of each photon lie on either side of the established thresholds. To confirm that a transition has indeed occurred, we check that the time delays for the next three photons after the detected transition correspond to the expected qubit state. If the time delay between any of the three consecutive photons appears to be between the two thresholds, the following dark or bright qubit state period is also excluded from the analysis. Time periods corresponding to the dark and the bright ion state were then evenly divided into time intervals equal to the desired integration time. Only periods that contain multiple integration time intervals were used in the analysis. During the time interval selection we specified that the neighbouring ions must be in the bright state to maximize the negative effect of the optical crosstalk and estimate the upper bound of the qubit state detection error.

After selecting the time intervals of the dark and bright states, we plotted the histograms for different integration times and obtained corresponding PDFs. In Figure~\ref{fig:epsart2}, the bright/dark state histograms for the second ion from the left are plotted for 5~ms and 30~ms integration times. Only the photons detected within the 9$\times$9 pixels square region of interest (ROI) were used. The size of the ROI was chosen to maximize the photon counts while minimizing the optical crosstalk.  For each integration time, at least 5$\times$10$^{4}$ time intervals were used for each ion.

There is a small but non-zero probability of spontaneous decay from the $5D_{5/2}$ dark state to the bright state during the integration time. This probability increases with increasing integration time, which could affect the overall detection fidelity if the photon number discrimination method is used. In presence of decay the dark state PDF becomes:
    
\begin{equation}
    \label{eqn:mod_poisson}
    p_{d} = \frac{\tau-t_{int}}{\tau}P(n,\bar{n}_d) +
    \frac{t_{int}}{\tau}\frac{\Gamma(\bar{n}_b,n+1)-\Gamma(\bar{n}_d,n+1)}{\bar{n}_b - \bar{n}_d},
\end{equation}

\noindent where $n$ is the number of photons, $P(n,\bar{n}_d)$ is the unperturbed Poisson distribution, $\bar{n}_b$ and $\bar{n}_b$ are the average numbers of detected photons, $\Gamma(\bar{n}_{d,b}, n+1)$ is the gamma distribution function, and $t_{int}$ is the integration time. For the 30~ms integration time the decay probability is 0.00096, which would lead to an additional error of $\sim4\times10^{-6}$ to incorrectly identify the dark state as bright. Since we selected pure bright/dark state intervals, the possibility of spontaneous emission occurring during the integration time is excluded in our estimation of the bright/dark state discrimination error. We separately calculate the qubit state detection error due to spontaneous emission and add it to the bright/dark state discrimination error.

There are multiple possible sources of the erroneous photon counts for an ion in the dark state, including the laser scattered from the trap surface, the intensifier dark counts, and optical/instrumental crosstalk. We found that the influence from the first two sources was negligible since the spatially uniform background/dark count rate was below 1~count/s within the ROI. Optical crosstalk was significant, with approximately 5.5\% of the fluorescence of a bright ion falling into the ROI of the neighboring ion in the chain. Due to the astigmatism of the optical system, there is a slight variation of the crosstalk due to the left neighboring ion compared to the right neighboring ion for each ion in the chain. This crosstalk leads to an increase in the average number of photon counts for the dark state histograms when  the neighboring ions are in the bright state. Optical crosstalk leads to the broadening of the dark state histogram and increases the detection error. We estimate that in the case of the diffraction-limited optics, the corresponding light leakage would be 0.4\%. Optical crosstalk can be significantly reduced by using optics with higher numerical aperture; for example, in~\cite{Crain19} a diffraction-limited objective lens with numerical aperture of 0.6 was used. A higher numerical aperture optics both reduces the size of the Airy patterns leading to a lower optical crosstalk, and increases the total collection efficiency.

\begin{figure}[!htb]
    \centering
        \includegraphics[width=0.45\linewidth]{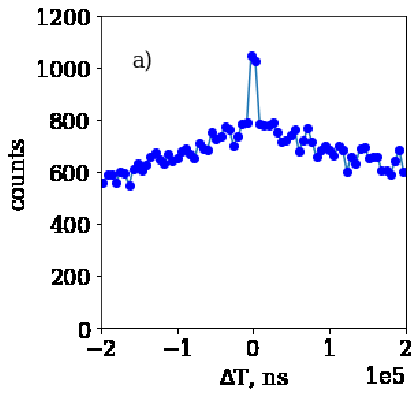}
        \includegraphics[width=0.45\linewidth]{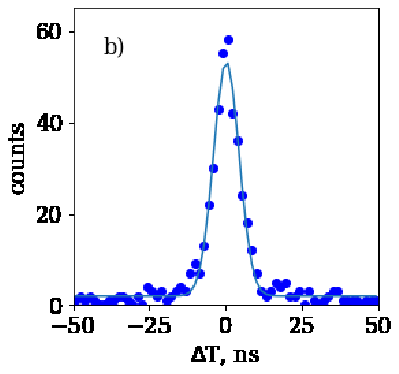}
	\caption{\footnotesize{Time difference distribution for two time ranges, $\pm 0.2$~ms (a) and $\pm 50$~ns (b). The peak at $\Delta \mathrm{T}=0$, which corresponds to the MCP afterpulses, is fit to a Gaussian with a width $\sigma = 4.2$~ns.
	}}
	\label{fig:dt}
\end{figure}

The instrumental crosstalk is caused by MCP afterpulsing in the intensifier of the camera. Electron avalanches in the MCP could result in the secondary electrons or ions producing independent hits in the vicinity of the primary hit \cite{Orlov2018, Orlov2019}. The time difference between the main hit and the afterpulse is small, so we can easily identify these cases as pairs of photons detected at the same time, looking at the time delay between photons detected from two neighboring bright ions. Figure~\ref{fig:dt} shows the time difference distributions between such photon detection events for two different time ranges.

The probability of detecting a fake hit due to the MCP afterpulses of 0.15\% was determined from the data by estimating the number of events in the peak at $\Delta \mathrm{T}=0$ and normalizing it to the total number of registered photons. 
Since all the detected photons have time-stamps, this source of crosstalk can be removed by ignoring hits at the dark ion location in a 50~ns window around the time when another photon was registered at the neighboring bright ion location. In our case, the contribution of this crosstalk source is very small, at the level of only about 0.7 photons/s on average, and we did not apply this post-selection in our analysis.

\begin{figure}
\includegraphics[width=3in]{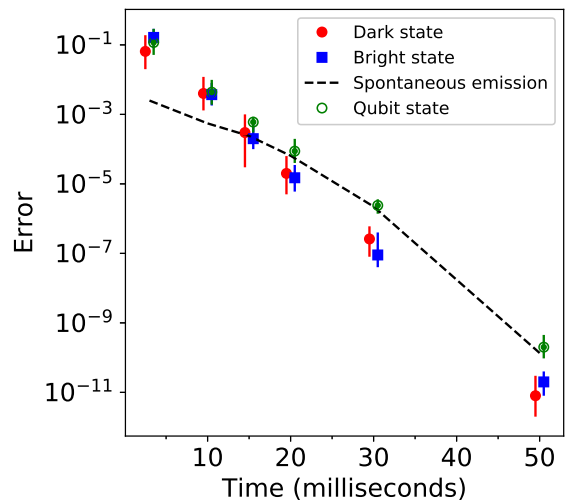}
\caption{\label{fig:epsart3} \footnotesize{Detection errors as a
function of integration time. Solid squares (circles) represent the data for the bright (dark) state discrimination errors. The optimal threshold value $n_{tr}$ is set by calculating where two Poisson curves corresponding to dark/bright ion states cross by extrapolating the curves, and rounding this value to an integer. The error bars are calculated by setting the threshold to $n_{tr}\pm$1. Horizontal offset of $\pm$ 0.5~ms between dark and bright detection errors was introduced for clarity. The dashed line is the error due to the $5D_{5/2}$ spontaneous decay, calculated using Eq.~1. The combined error, shown in open circles, is the qubit state detection error that takes into account the bright/dark state discrimination error and the error due to spontaneous emission.}}
\label{fig:error}
\end{figure}

The summary of the single qubit state detection error for the second ion from the left is plotted as a function of the integration time in Figure~\ref{fig:error}. The data follow the expected trend for discriminating two random variables with Poisson distributions. The dark/bright state errors are somewhat different due to the choice of $n_{tr}$. The average of the two errors is used to calculate the qubit state detection error for each ion. The fidelity of detection of the N-qubit state is calculated as a product of the single qubit fidelities $\Pi(1-\epsilon_{i})$, where $\epsilon_{i}$ is the detection error of the $i$-th ion. 

The average bright/dark state discrimination error at 30~ms integration time varies between 7(6)$\times10^{-9}$ for the outer two ions, where the optical crosstalk is lower, and 5.8(3.8)$\times10^{-7}$ for the inner ions. The additional error due to the qubit spontaneous emission is 5.4(0.4)$\times10^{-6}$ for the outer qubits and 3(1)$\times10^{-6}$ for the inner ones, the difference being due to the different values of $n_{tr}$. The single qubit state detection error averaged over all four qubits is 4.2(1.5) ppm, and the total detection error for the 4-qubit state is 17(2) ppm.

In summary, we demonstrated simultaneous detection of four $^{138}$Ba$^+$ ion qubits in a linear chain achieving a qubit state detection error of 4.2(1.5) ppm for a single ion in the presence of bright neighbouring ions with a 30~ms detection time, considerably improving previous results. The detection  error of the four-qubit state was 17(2) ppm. The qubit state detection fidelity is  limited by the lifetime of the $5D_{5/2}$ qubit state, which for barium ion is approximately 32~s, by far the longest of all ion qubit candidates, making it suitable for the highest qubit state detection fidelity. Further reduction of the detection error can be achieved by improving the collection efficiency of the optical system and reducing the crosstalk between neighboring ions.  We conclude that the fast time-stamping camera used in the experiments offers a straightforward route for scaling up the number of simultaneously detected qubits in a linear ion chain. It can be increased to about 30 qubits in the present linear configuration with 10-pixel spacing between the ions, and to 60 qubits with reduced optical magnification giving a 5-pixel ion spacing. In the two-dimensional trap setup \cite{IvoryM}, the number of simultaneously detected qubits can easily be a few hundred. Even with a few kHz photon detection rate per ion, the total photon rate will still be below the maximum allowed rate of about 10$^7$ photons/s. The camera data can be promptly analysed in real time providing input for the error correction algorithms.

We thank Michael Keach, Maverick J. Millican and Kurt A. Delegard for assistance with the datasets, analysis and optical setup. This work was supported by the U.S. Department of Energy QuantISED award, U.S. National Science Foundation award PHY-2011503 and by the grant LM2018109 of Ministry of Education, Youth and Sports as well as by Centre of Advanced Applied Sciences CZ.02.1.01/0.0/0.0/16-019/0000778, co-financed by the European Union.

\bibliography{apssamp}

\end{document}